\def\rfr#1{eq. (\ref{#1})}

\def\bar{\begin{eqnarray}}
\def\ear{\end{eqnarray}}
\def\bb{\bibitem}
\def\eqi{\begin{equation}}
\def\eqf{\end{equation}}
\def\eqia{\begin{eqnarray}}
\def\eqfa{\end{eqnarray}}
\def\rp#1#2{{#1\over#2}}

\def\lb#1{\label{#1}}

\def\oc2{$\mathcal{O}(c^{-2})$}

 \documentclass[11pt]{article}
\usepackage{amsmath,amsthm,amscd,amssymb}
\usepackage{latexsym}
\usepackage{graphicx,epsfig}

\begin{document}

\noindent{\bf \LARGE{Jupiter, Saturn and the Pioneer anomaly: a planetary-based independent test}}
\\
\\
\\
{Lorenzo Iorio}\\
{\it Viale Unit$\grave{a}$ di Italia 68, 70125\\Bari, Italy
\\tel./fax 0039 080 5443144
\\e-mail: lorenzo.iorio@libero.it}

\begin{abstract}
In this paper we use the ratio of the corrections to the standard Newtonian/Einsteinian secular precessions of the longitudes of perihelia of Jupiter and Saturn, recently estimated by the Russian astronomer E.V. Pitjeva by fitting almost one century of data with the EPM ephemerides, to make an independent, planetary-based test of the hypothesis that the Pioneer anomaly (PA), as it is presently known in the 5-10 AU region, is of gravitational origin. Accounting for the errors in the determined apsidal extra-rates and in the values of the PA acceleration at the orbits of Jupiter and Saturn the answer is negative. If and when the re-analysis of the entire Pioneer 10/11 will be completed more firm conclusions could be reached. Moreover, it would also be important that  other teams of astronomers estimate independently their own corrections to the perihelion precessions.
\end{abstract}


 \section{Introduction}
The so-called Pioneer Anomaly \cite{And98,And02} (PA in the following) consists of an unmodeled, small, constant Doppler blueshift of the Pioneer 10/11 spacecraft occurred at heliocentric distances of about 20-70 AU which can be interpreted as a constant acceleration $A_{\rm Pio}=(8.74\pm 1.33)\times 10^{-10}$ m s$^{-2}$ directed approximately towards the Sun. Subsequent independent studies \cite{Mar02,Ols07} confirmed the existence of such an effect in the Pioneer 10/11 data.

In \cite{Nie05,Nie07} an analysis of early-and poorly modeled-data of Pioneer 11  has showed that it may have been an onset of PA between about the Jupiter ($a=5.2$ AU) and Saturn ($a=9.5$ AU) encounters. In Figure 4 of \cite{Nie05}, reproduced, e.g., as Figure 2 in \cite{Nie07} and appeared first in \cite{memos}, it can be noted that
\begin{equation}\left\{\begin{array}{lll}
A_{\rm Pio}^{(\rm Jup)}\simeq (0.8\pm 1.4)\times 10^{-10}\ {\rm m\ s}^{-2},\\\\
A_{\rm Pio}^{(\rm Sat)}\simeq (1.8\pm 6.4)\times 10^{-10}\ {\rm m\ s}^{-2}.
\lb{jupsatacc}\end{array}\right.\end{equation}
The author of \cite{Nie07} propose to use the 2007-2008 data of the ongoing New Horizons (http://pluto.jhuapl.edu/) mission to the Pluto system and the Kuiper Belt
to make a spacecraft-based independent test of  PA in the\footnote{The New Horizons spacecraft took a gravity assist by Jupiter in February 2007 and should reach the orbit of Saturn in mid-2008.} 5-10 AU region.

Soon after its discovery, a seemingly un-endless flood of papers looking for non-conventional, exotic explanations of PA started: see, e.g., \cite{Ber04,Izz06,pallee} and references therein. The hypothesis that PA is due to some long-range modifications of the known Newtonian/Einsteinian laws of gravity, and not due to some technical issues peculiar to the twin spacecraft (see \cite{pallee} and references therein), has been recently put on the test in a phenomenological and independent way by analyzing the orbital motions of the planets\footnote{In \cite{And02} the inner planets were considered, but this is of little interest because PA manifested in the outer regions of the Solar System. Attempts to investigate the consequences of a PA-like extra-acceleration on  long-period comets can be found in \cite{And02b}. The authors of \cite{Wal07} used a sample of trans-Neptunian objects between 20 and 100 AU.} of the Solar System \cite{Ior06,Ior07,Tan07} orbiting in the regions in which PA manifested itself in its presently known form. Indeed, if it is of gravitational origin it must fulfil the equivalence principle and also such natural bodies   must be affected by an extra-acceleration with the physical characteristic of that allegedly inducing PA. The outcome of \cite{Ior06,Ior07,Tan07,Wal07}, performed with different and independent approaches, is negative: no effects which could be due to a PA-type anomalous acceleration can be detected in the motions of natural bodies moving at $20<r<100$ AU. In this paper we wish to extend such a strategy to the 5-10 AU onset region by looking at the orbital motions of Jupiter and Saturn.
\section{Anomalous perihelion precessions of Jupiter and Saturn and their (in)compatibility with a PA onset}
In \cite{Ior06,San06} it was shown that a small uniform, constant  and radial acceleration $A$ like the one which should be responsible of PA would induce a secular, i.e. averaged over one orbital period, precession of the longitude of pericentre $\varpi$ of a test particle
\eqi\left<\dot\varpi\right>=A\sqrt{\rp{a(1-e^2)}{GM}},\lb{rat}\eqf where $G$ is the Newtonian gravitational constant, $M$ is the mass of the central body orbited by the test particle, $a$ and $e$ are the semimajor axis and the eccentricity, respectively, of its orbit.

The Russian astronomer E.V. Pitjeva   (Institute of Applied Astronomy, Russian Academy of Sciences) recently processed almost one century of planetary data of different kinds with the dynamical models of the EPM2004 ephemerides \cite{Pit05a}  which include almost\footnote{With the exception of the general relativistic Lense-Thirring effect and of the classical force exerted by the Kuiper Belt Objects (KBOs).} all Newtonian and Einsteinian known forces. Among the solve-for parameters estimated with the least-square approach she  also determined corrections $\Delta\left<\dot\varpi\right>$ to the classical/relativistic secular apsidal precessions of the inner \cite{Pit05b} and of some of the outer \cite{Pit06} planets of the Solar System: such corrections, by construction, should account for any unmodelled force existing in nature being, thus, well suited for our purposes.   In the case of Jupiter and Saturn, for which only optical data were used apart from some radiotechnical data for Jupiter \cite{Pit05a}, we have  \cite{Pit06}
\begin{equation}\left\{\begin{array}{lll}
\Delta\left<\dot\varpi\right>^{(\rm Jup)}=0.0062 \pm 0.036\ {\rm arcsec\ cy}^{-1},\\\\
\Delta\left<\dot\varpi\right>^{(\rm Sat)}=-0.92\pm 2.9\ {\rm arcsec\ cy}^{-1};
\lb{jupsatprec}\end{array}\right.\end{equation}
the errors released here have been obtained by re-scaling the formal, statistical ones \cite{Pit06} by a factor 10 in order to get realistic estimates for them.

The determined apsidal extra-rates for Jupiter and Saturn of \rfr{jupsatprec} can now be compared to the predicted anomalous precessions of \rfr{rat} evaluated for \rfr{jupsatacc}; note that, in view of the small eccentricities of the orbits of Jupiter ($e=0.0483$) and Saturn ($e=0.0541$), it is certainly quite reasonable assuming $A_{\rm Pio}$ constant over the orbital paths of each planet so that the use of \rfr{rat} is allowed, although $A_{\rm Pio}$  differs from one planet to the other according to \rfr{jupsatacc}. Thus, we have
\begin{equation}\left\{\begin{array}{lll}
\left<\dot\varpi\right>_{\rm Pio}^{(\rm Jup)}=4\pm 7\ {\rm arcsec\ cy}^{-1},\\\\
\left<\dot\varpi\right>_{\rm Pio}^{(\rm Sat)}=12\pm 43\ {\rm arcsec\ cy}^{-1};
\lb{jupsatprecpio}\end{array}\right.\end{equation}
It can be noted that anomalous perihelion precessions as large as predicted by \rfr{jupsatprecpio} are compatible\footnote{As shown in \cite{Ior07}, it is not so for the apsidal extra-rates predicted for Jupiter and Saturn by the Yukawa-like model used in \cite{Bro06} to fit the entire Pioneer 10/11 data set. Indeed, the errors in the parameters entering the Yukawa-type model of \cite{Bro06} are smaller than their best estimates, so that well definite non-zero extra-precessions of perihelion are predicted.} with the determined extra-rates of \rfr{jupsatprec} (in the sense that they are both compatible with zero) if we consider each planet one at a time.   But we are able to perform a tighter test by taking the ratio of the estimated extra-rates of \rfr{jupsatprec} and comparing it to the ratio of the PA-type precessions of \rfr{rat} evaluated with \rfr{jupsatacc}. Indeed, we have
\eqi \left|   \rp{\Delta\left<\dot\varpi\right>^{\rm (Jup)}A_{\rm Pio}^{\rm (Sat)}}{\Delta\left<\dot\varpi\right>^{\rm (Sat)}A_{\rm Pio}^{\rm (Jup)}} -\sqrt{ \rp{a_{\rm Jup}(1-e_{\rm Jup}^2)}{a_{\rm Sat}(1-e_{\rm Sat}^2)} } \right|=0.75\pm 0.05,\lb{spiff}\eqf
which is not compatible with zero as, instead, it should be if the estimated corrections to the perihelion precessions of Jupiter and Saturn of \rfr{jupsatprec} were due to an anomalous PA-type acceleration having the values of \rfr{jupsatacc}. The uncertainty in \rfr{spiff} has been conservatively computed by linearly summing the bias due to the errors in \rfr{jupsatacc} and \rfr{jupsatprec}. Note that our conclusion would not change even if we re-scaled the formal error in the estimated apsidal extra-precession of Saturn-for which only optical data \cite{Pit05a} were used-by a factor 100.
\section{Conclusions}
In this paper we investigated the possibility that an onset of PA occurred in the 5-10 AU region by assuming a gravitational origin for the anomalous behavior of the Pioneer 10/11 spacecraft. In this case, unless postulating a violation of the equivalence principle, the putative exotic force responsible for PA must also act upon the planets of the Solar System orbiting the spatial regions in which PA manifested itself in its presently known form. By taking the ratio of the corrections to the usual Newtonian/Einsteinian secular precessions of the perihelia of Jupiter and Saturn recently estimated by E.V. Pitjeva with the EPM ephemerides it turns out that the possibility that an extra-force with the same physical characteristics observed so far in the 5-10 AU region for PA affects Jupiter and Saturn as well is ruled out. In obtaining such a conclusion we re-scaled by a factor 10 the formal errors in the planetary apsidal extra-rates used. If and when other teams of astronomers will independently determine their own anomalous precessions of perihelia, and if and when the re-analysis of the entire Pioneer10/11 data set will be completed \cite{let1,let2}   more firm conclusions could be traced.


\end{document}